\documentclass[a4paper,11pt]{article}
\usepackage{pos}
\usepackage{bm}
\usepackage{xspace}
\usepackage{setspace}
\usepackage{enumitem}

\newcommand{\tev}{\,TeV\xspace}

\title{Detecting and characterizing pulsar halos with the Cherenkov Telescope Array Observatory}
 \ShortTitle{PSR halos with CTA}

\author*[a, b]{Christopher Eckner}
\forColl{CTA Consortium} 

\affiliation[a]{LAPTh, CNRS, F-74000 Annecy, France}
\affiliation[b]{LAPP, CNRS, F-74000 Annecy, France}

\emailAdd{eckner@lapth.cnrs.fr}

\abstract{The recently identified source class of pulsar halos may be populated and bright enough at TeV energies to constitute a large fraction of the sources that will be observed with the Cherenkov Telescope Array (CTA), especially in the context of the planned Galactic Plane Survey (GPS). In this study, we examine the prospects offered by CTA for the detection and characterization of such objects. CTA will cover energies from 20 GeV to 300 TeV, bridging the ranges already probed with the Fermi Large Area Telescope and High Altitude Water Cherenkov Observatory, and will also have a better angular resolution than the latter instruments, thus providing a complementary look at the phenomenon. From simple models for individual pulsar halos and their population in the Milky Way, we examine under which conditions such sources can be detected and studied from the GPS observations. In the framework of a full spatial-spectral likelihood analysis, using the most recent estimates for the instrument response function and prototypes for the science tools, we derive the spectral and morphological sensitivity of the CTA GPS to the specific intensity distribution of pulsar halos. From these, we quantify the physical parameters for which pulsar halos can be detected, identified, and characterized, and what fraction of the Galactic population could be accessible. We also discuss the effect of interstellar emission and data analysis systematics on these prospects.}

\ConferenceLogo{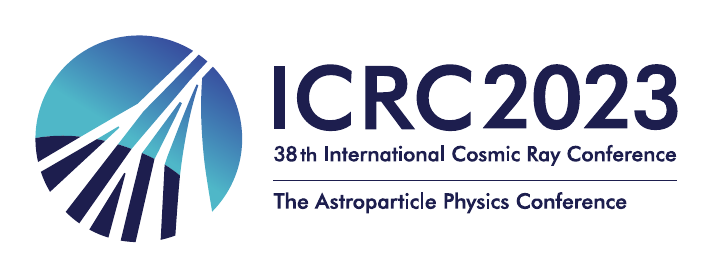}

\FullConference{%
38th International Cosmic Ray Conference (ICRC2023)\\
  26 July - 3 August, 2023\\
  Nagoya, Japan}

\begin{document}
\maketitle

\section{Introduction}
\label{intro}

Pulsars -- high-energy astrophysical objects born in violent supernova explosions -- have claimed a prominent position in contemporary astroparticle physics in the past decade. Not only are their emissions observable over a broad range of the electromagnetic wave band, but they may also be one of the main contributors to the observed local flux of cosmic-ray leptons \cite{Evoli:2021}. The \textit{Fermi}-LAT has detected over 270 pulsars due to their gamma-ray emission at GeV energies, far exceeding detections in the pre-Fermi era \cite{Thompson:2008, latpulsars:2022}. Ground-based Imaging Atmospheric Cherenkov Telescopes (IACTs), such as H.E.S.S., advanced our understanding of the TeV emission of pulsars during the later stages of their evolution, i.e.~when they have formed so-called pulsar wind nebulae (PWNe) \cite{Abdalla:2018a,Abdalla:2018b}. The gamma-ray emission patterns associated with PWNe are influenced by the pulsar's natal kick and the dynamics of the parent supernova remnant. They become complex as pulsar, PWNe and supernova remnant evolve \cite{Abdalla:2019,Principe:2020}. Observations with the High-Altitude Water Cherenkov observatory (HAWC) revealed extended gamma-ray emission, initially referred to as TeV halos \cite{Linden:2017} but more recently replaced by the term \emph{pulsar halos}, around two middle-aged pulsars -- PSR J0633+1746 (Geminga) and PSR B0656+14 (Monogem) --, providing new insights into particle acceleration and escape mechanisms \cite{Abeysekara:2017b}. The pulsar halos' emission appears to be fueled by the central pulsar since, in those two cases, the central pulsar has left the region dynamically dominated by the associated PWN. Interestingly, it is possible to describe the observed pulsar halo emission with a phenomenological diffusion model featuring a suppressed diffusion coefficient with respect to the values inferred in the remainder of the Milky Way. The cause of the inhibited diffusion is not well understood and subject of ongoing research \cite{LopezCoto:2018,Fang:2019,Mukhopadhyay:2021}. Despite these observations, the commonness of pulsar halos in the Galaxy is unknown. Upcoming observations with HAWC, LHAASO, H.E.S.S., and the Cherenkov Telescope Array (CTA) are expected to contribute significantly to this field \cite{Lopez-Coto:2022igd, Abdalla:2021, Acharya:2019}. This work -- whose unabridged content is published at \cite{Eckner:2022gil} -- aims to use the upcoming CTA Galactic Plane Survey (GPS) \cite{Remy:2022} to constrain current phenomenological pulsar halo models. We will determine the survey's sensitivity to the $\gamma$-ray spectrum and morphology of individual halos, and estimate the fraction of the population detectable by CTA.

\section{Sky model}
\label{model}

This study focuses on the emission components considered in our simulations and analyses of CTA observations. These components include the phenomenological halo model, a synthetic halo population for the entire Galaxy, and various models for the astrophysical background components. The halo model has been utilized to predict the detectability of individual halos, while the synthetic halo population is used to determine the potential detectability fraction within the context of CTA's GPS observations \cite{Martin:2022,Tang:2019}.

\noindent\textbf{Individual pulsar halo model.} The individual halo model, introduced by \cite{Martin:2022}, simulates the behavior of electron-positron pairs accelerated and released into the surrounding medium by a pulsar shortly after its birth. These particles diffuse isotropically in a medium with a two-zone structure representing different diffusion properties. Particles lose energy and radiate via synchrotron and inverse-Compton scattering processes in assumed typical magnetic and radiation fields of the Interstellar Medium (ISM) \cite{Tang:2019}. The reference model is a pulsar with a current age of 200 kyr and a spin-down power of $10^{35}$ erg/s, with particle injection assumed to start 60 kyr after birth \cite{Zhang:2020a,Martin:2022}. The injected particles have a broken power-law spectrum up to a cutoff at 1 PeV. Diffusion suppression is by a factor 500 within a diffusion zone region of $r_{\mathrm{diff}} = 50$ pc radius.

\noindent\textbf{Pulsar halo population model.} The model setups mentioned above are used to ascertain the conditions for the potential observation of a typical halo in the CTA GPS \cite{Martin:2022b}. Using a synthetic population of young pulsars, we estimate the number of objects potentially accessible to the survey. However, the detectability criteria are only strictly valid for one set of halo model parameters, therefore, the prospects at the population level should be taken as approximations.

\noindent\textbf{Large-scale diffuse backgrounds.} In our analyses, we included a model for large-scale interstellar emission (IE) from the Galactic population of cosmic rays interacting with the ISM. While this component has been well mapped at GeV energies with the \textit{Fermi}-LAT, it is less established in the TeV range \cite{Fermi-LAT:2012edv}. To model this, we refer to a recent study \cite{Luque:2022buq} based on available GeV to PeV gamma-ray data and local, charged cosmic ray measurements. In terms of consistency, we also consider a model adopted in a reference prospect study of the CTA GPS \cite{Dundovic:2021ryb}.

\section{Data Simulations and Analysis}\label{data}

\subsection{Survey Simulation}\label{data:sim}

The upcoming CTA GPS promises extensive coverage of the Galactic Plane at TeV energies, comprising short and long-term programs totalling 1620 hours of observation over a decade \citep{Acharya:2019}. Our simulation follows the strategy outlined by \cite{Remy:2022}, ensuring optimal instrument performance by minimizing Moon light contamination and zenith angle. The survey's varying exposure is divided into five distinct segments: Inner Galaxy, Cygnus/Perseus, Anticenter, and two chunks for Vela/Carina \citep{Acharya:2019}.

The Inner Galaxy, the region with the youngest pulsars and thus pulsar halos, receives the most extensive exposure. Our reference coordinates for this study are $(l,b) = (-10^{\circ}, 0^{\circ})$ within this region, despite it being a conservative choice given the intensity of IE.

\begin{figure}
\centering
\includegraphics[width=0.5\textwidth]{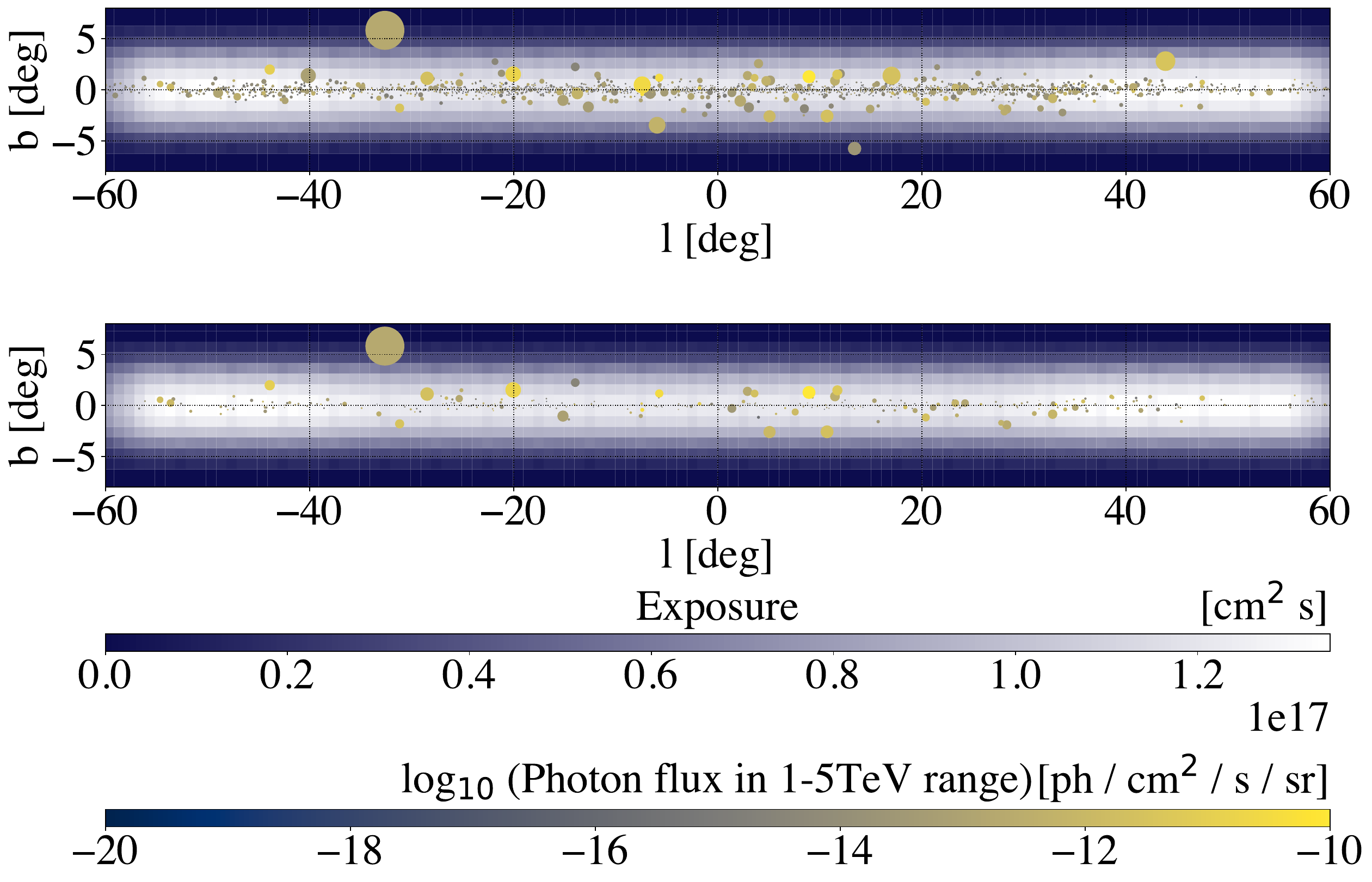}
\caption{A synthetic pulsar halo population overlaid on an exposure map for the central regions surveyed in the GPS.  \label{fig:GPSexp}}
\end{figure}

We use the latest instrument response functions (IRFs) labeled \texttt{prod5} for the different configurations at the North and South sites. These files also quantify the expected irreducible instrumental background (CR) of CTA due to charged cosmic rays misclassified as gamma-ray events. The data files are publicly available at \citep{cherenkov_telescope_array_observatory_2021_5499840}. We consider the initial "alpha" configuration to reflect the real number of telescopes given the budget. We use the optimized event reconstruction quality and background cuts derived from Monte Carlo simulations of 50-hour observations. In Fig.~\ref{fig:GPSexp} we display the exposure achieved with the GPS in the inner Galaxy and overlay this map with the pulsar halo population considered in this work.

\subsection{Analytical Framework}
\label{data:ana}

To analyze the potential for detecting and studying pulsar halos with CTA, we apply the statistical inference framework of \cite{CTA:2020qlo}. We employ a three-dimensional likelihood analysis for binned data and Poisson statistics. Our region of interest (ROI) is centered on $(l,b) = (-10^{\circ}, 0^{\circ})$ with a size of $6^{\circ} \times 6^{\circ}$ and pixel size of $0.02^{\circ} \times 0.02^{\circ}$. The energy range is from 0.1 to 100\tev, binned logarithmically with 15 bins per decade.

Our signal model in the ROI is a linear combination of templates representing the expected instrumental background $\{B_l\}_{l\in L}$ and astrophysical signal components $\{S_k\}_{k\in K}$. The likelihood function is given by:
\begin{equation}
\mathcal{L}\!\left(\left.\bm{\mu}\right|\bm{n}\right)=\prod_{i,j} \frac{\mu_{ij}^{n_{ij}}}{\left(n_{ij}\right)!}e^{-\mu_{ij}}\mathrm{,}
\end{equation}
where $\bm{\mu}$ represents the model used to fit $\bm{n}$, the CTA experimental data (mock data).

Our model is generally expressed as:
\begin{equation}
\bm{\mu}= \sum_{k\in K} \bm{S}_k(\theta^S_k) + \sum_{l\in L} \bm{B}_l (\theta^B_l) \mathrm{,}
\end{equation}
where $\{\theta^S_k\}_{k \in K}$ and $\{\theta^B_l\}_{l \in L}$ represent multiplicative parameters adjusting the spectral and angular dependence of the signal and instrumental background templates. For simplicity, we often consider a model with a global renormalization parameter of predefined templates.

The capability to distinguish between alternative hypotheses regarding the measured signal is assessed using the log-likelihood ratio test statistic (TS). The TS is given by:
\begin{equation}
\label{eq:tsdet}
\textrm{TS} = 2 \left( \ln \left[ \frac{\mathcal{L} \left( \bm{\mu} ( { \hat{\theta}^{S_{\rm test}}k }, { \hat{\theta}^B_l }) | \bm{n} \right)}{\mathcal{L} \left( \bm{\mu} ({ \hat{\theta}^{S_{\rm null}}_k }, { \hat{\theta}^B_l }) | \bm{n} \right)} \right] \right) \mathrm{,}
\end{equation}
where the hatted quantities refer to the best-fit values for all model parameters.

\noindent\textbf{Estimation of Detectability.} The detectability of astrophysical components can be quantified using the test statistic (TS) given by Eq.~\ref{eq:tsdet} when assuming that in the null hypothesis $\hat{\theta}_k^{S_{\rm null}}\equiv 0$. Under relatively mild conditions satisfied here, the TS distribution follows a non-central chi-square distribution with $K$ degrees of freedom. In most cases, we examine a single signal template, which simplifies the problem to a half-chi-square distribution with a single degree of freedom. Signal detection at the 5-sigma level corresponds to a TS value of approximately 25.

\noindent\textbf{Spectral Sensitivity Analysis.} We evaluated both spectral sensitivity and model-independent angular sensitivity. As concerns the spectral sensitivity, we use the entire energy range and derive the sensitivity energy bin by energy bin. This sensitivity measures the flux level that results in detection with a TS of 25 and at least ten detected gamma-ray events in each energy bin. The process involves preparing mock datasets featuring the CR background as well as IE and the pulsar halo source at a given normalization. We fit the full model and background-only model (CR + IE) until we find the value of the pulsar halo normalization that yields a TS of 25 over the null hypothesis.

\noindent\textbf{Angular Decomposition of a Pulsar Halo Signal.} We decomposed individual pulsar halos using a step-by-step fitting process of growing concentric annuli obtained from the original pulsar halo model. We evaluated Eq.~\ref{eq:tsdet} iteratively, either expanding the annulus width or adding a new annulus to the model based on the resulting TS value. This procedure continued until no significant
signal was observed. Finally, the parameters and errors for all significant annuli were considered as the recovered angular decomposition of the input signal.

\noindent\textbf{Handling of Systematic Errors.} In order to deal with systematic uncertainties, we used an effective Poisson likelihood function \cite{Silverwood:2014yza, CTA:2020qlo}. This function incorporates nuisance parameters per pixel and energy bin of the binned model, which mimic Gaussian noise affecting each pixel independently. However, this method only accounts for uncertainties that linearly affect the number of expected gamma rays, such as CTA's effective area. Systematic errors were integrated into our sensitivity derivations by adding these nuisance parameters to the set of background parameters $\{\theta^B_l\}_{l \in L}$.

\section{CTA's sensitivity to pulsar halos}
\label{sec: results}


\subsection{Exploring the sensitivity to the phenomenological pulsar halo model}\label{sec:sens_psrhalo}

\begin{figure}
\centering
\includegraphics[width=0.49\textwidth]{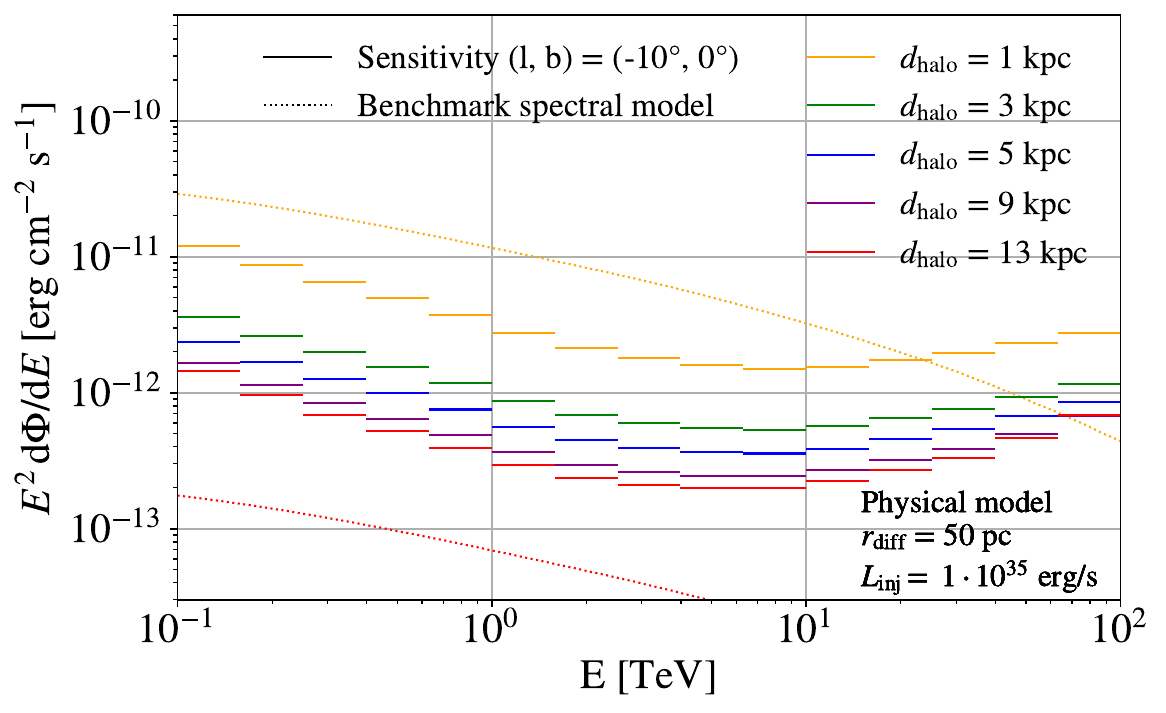}
\hfill\includegraphics[width=0.49\textwidth]{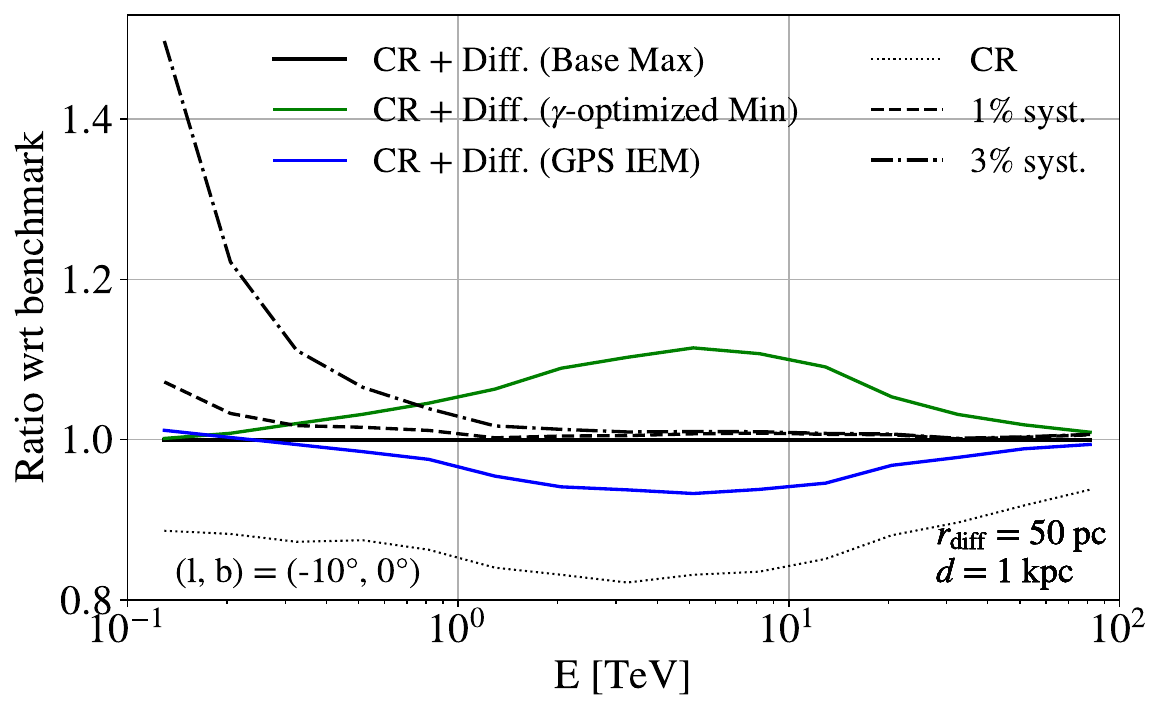}
\caption{(\emph{Left}:) Differential spectral energy sensitivity to a baseline halo model, positioned at different distances from the observer. Results are shown for analysis focusing on a $6^{\circ}$ region centred at $(\ell, b) = (-10^{\circ}, 0^{\circ})$, and are overlaid with the halo model intensities for halos at 1 and 13 kpc distance in orange and red dashed lines, respectively. (\emph{Right}:) Differential sensitivity to our baseline halo model at 1 kpc distance considering different diffuse emission models with respect to our benchmark background model (cosmic rays + Base Max IEM). The dashed and dash-dotted black lines show the benchmark model sensitivity with added systematic uncertainties of 1\% and 3\%, respectively. \label{fig:singlehalo}}
\end{figure}

\noindent\textbf{Spectral Sensitivity.} The spectral sensitivity, visualized in the left panel of Fig.~\ref{fig:singlehalo}, declines with an enlarged halo extension, a consequence of a closer positioning. For a more intuitive understanding, the sensitivities to the considered pulsar halo models align approximately with a Gaussian source of the corresponding angular size. The sensitivity to the physical pulsar halo model shows less energy variation due to the energy-dependent shrinking of the source size. This effect is primarily due to the increase in energy losses with increasing particle energy, offsetting the deteriorating instrument sensitivity beyond roughly 10 TeV. At larger distances, exceeding approximately 4-5 kpc, the sensitivity trend starts to flatten, as the halo's typical angular size becomes similar to the instrument's angular resolution. Meanwhile, the flux from a halo decreases inversely with the square of the distance. The flux levels of our reference individual halo model are portrayed in the left panel of Fig.~\ref{fig:singlehalo}. Spectral studies ranging from hundreds of GeV to tens of TeV seem feasible for nearby halos at distances of 1-3 kpc, involving injection powers a few times above that deduced for Geminga, or roughly $10^{35}$ erg/s.

\noindent{\bf Impact of the IE Model:} The right panel of Fig.~\ref{fig:singlehalo} explores the effect of three assumptions: the scenario without IE, the $\gamma$-optimized Min model \cite{Luque:2022buq}, and the IE model used in the GPS publication \cite{Remy:2022}. The IE impact is more significant for an extended source, though limited to a $20\%$ impact.

\noindent{\bf Systematic Uncertainty:} The analysis here relies heavily on accurate models for emission components. However, many instrumental effects, particularly those affecting smaller-scale field-of-view observations, are difficult to model but could have a significant influence. Following \cite{CTA:2020qlo}, we consider an impact scale of 0.1$^{\circ}$ and the impact magnitude of $1\%$ and $3\%$. The impact of these effectively modeled systematic uncertainties is largest at the lowest GeV energies considered in our study. This regime is already systematics-dominated due to the large amount of available gamma-ray events. The effect of these uncertainties is less than $10\%$ around 1 TeV, the energy range with the highest sensitivity to this kind of gamma-ray source class. 

\noindent\textbf{Angular Sensitivity:} In the left panel of Fig.~\ref{fig:halopop} we present the results regarding the angular decomposition of the baseline pulsar halo model with a diffusion zone size of 50 pc at a distance of 1 kpc from Earth. The experimental CTA mock data contains the halo source as well as CR background and IE (Base Max). These results highlight the potential of CTA measurements to study the angular profile of pulsar halos. In particular, the 0.1 - 1 TeV energy range allows us to probe the halo's morphology up to 30 pc from the central pulsar and, in addition, it is possible to distinguish between different parameters of the phenomenological pulsar halo model like the diffusion zone size $r_{\mathrm{diff}}$.

\subsection{Prospects for a Galactic pulsar halo population}
\label{sec:sens_psrpop}

\begin{figure}
\centering
\includegraphics[width=0.53\textwidth]{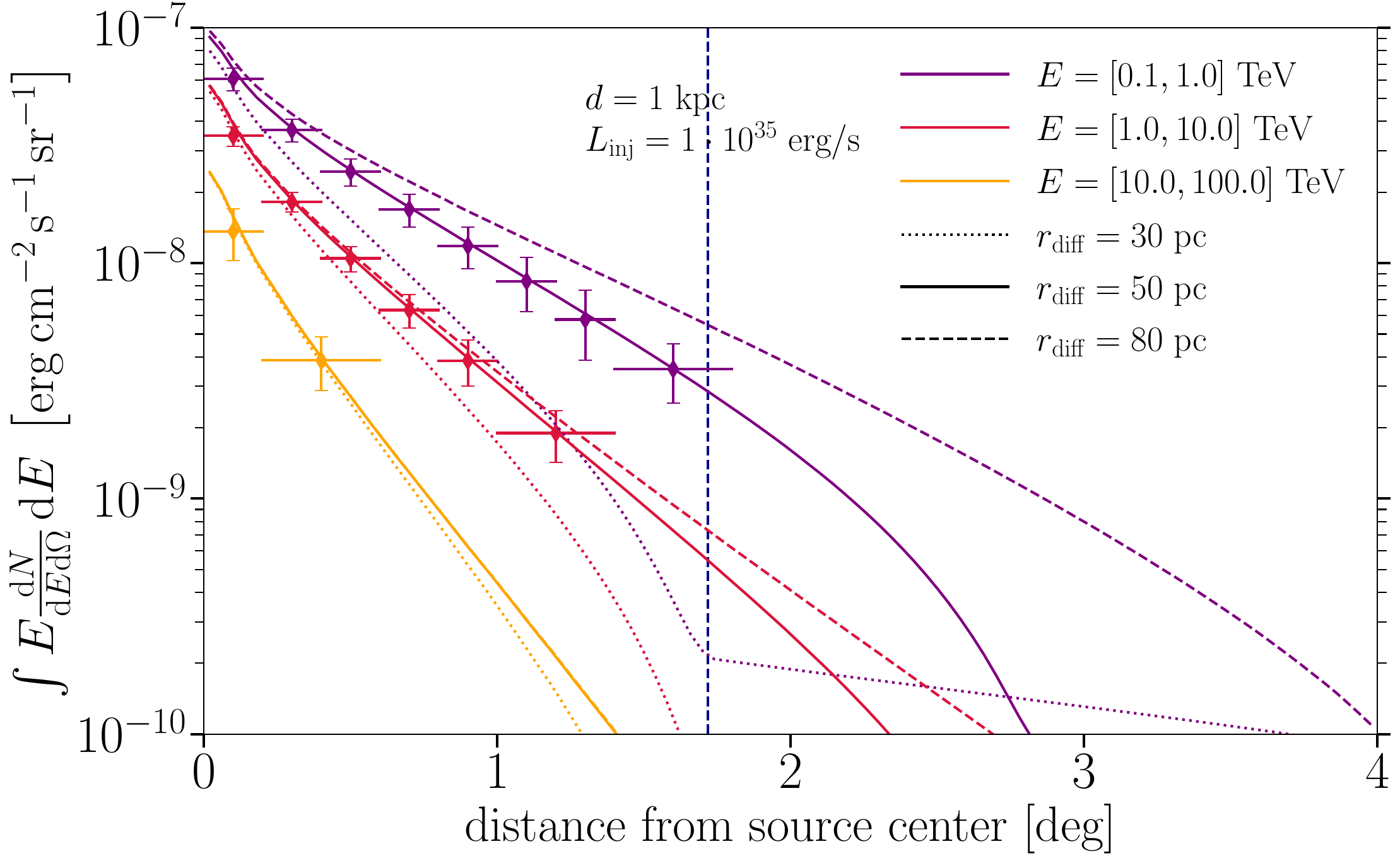}
\hfill\includegraphics[width=0.45\textwidth]{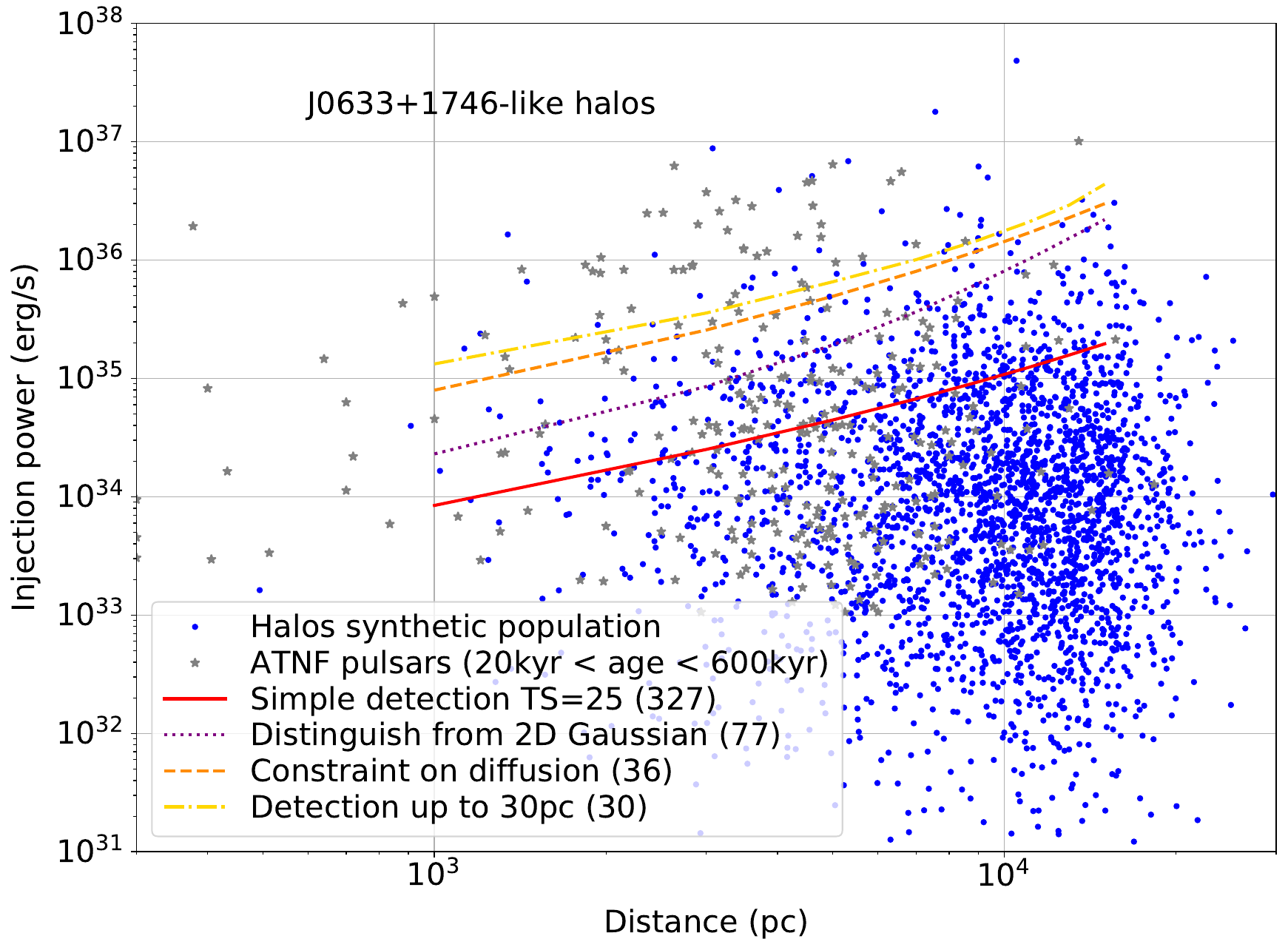}
\caption{(\emph{Left}:) Model-dependent angular sensitivity to a pulsar halo at 1 kpc distance and $r_{\mathrm{diff}} = 50$ pc studied under the conditions of the GPS, overlaid with the baseline halo model predictions for a diffusion zone size of 30/50/80 pc. The sensitivity analysis is decomposed into annuli with a width of $0.2^{\circ}$, and split into three energy bands. (\emph{Right}:) Sensitivity of the GPS in terms of particle injection power as a function of distance. For each sensitivity curve, the numbers in parentheses indicate the number of mock halos lying above the curve. Grey data points indicate the positions and injection powers of known pulsars taken from the ATNF database \url{https://heasarc.gsfc.nasa.gov/W3Browse/all/atnfpulsar.html}. \label{fig:halopop}}
\end{figure}

We estimate the fraction of the pulsar halo population that should be detectable with CTA, assuming Geminga-like diffusion properties and supposing that all middle-aged pulsars develop a halo. The results of this population study are shown in the right panel of Fig.~\ref{fig:halopop}. We assess different properties of the halo population that CTA is capable of analyzing:
\setlist{nolistsep}
\begin{enumerate}[noitemsep]
    \item Injection power such that the simulated halo signal is detected with a TS of 25 over the full energy range, using the true halo model in the fit process (red line).
    \item Injection power such that a fit of the simulated halo signal with the true halo model is significantly preferred over a simple energy-independent 2D Gaussian intensity distribution (dashed purple line).
    \item Injection power such that a fit of the simulated halo signal with the true halo model is significantly preferred over the true model truncated at 30 pc from the pulsar (dashed orange line).
    \item Injection power such that a fit of the simulated halo signal with the true halo model is significantly preferred over an alternative halo model having a 50\% higher suppressed diffusion coefficient (dashed yellow line). 
\end{enumerate}
In this setup, CTA is expected to detect around 300 pulsar halos of which more than 70 representatives will have a significant energy-dependent morphology. In addition, up to 30 of these objects can be spatially decomposed to study the diffusion conditions around the halo's central pulsar. These prospects are reduced by a factor of about four when assuming that all pulsar halos resemble the case of Monogem. However, it must be noted that it is not clear at all that the entirety of Galactic pulsars will develop halos. In such less optimistic scenarios \cite{Martin:2022}, the fraction of accessible pulsar halos may only be 5-10\% of the stated values. 

\section{Conclusions}
\label{sec:conclusion}

This study offers an evaluation of the potential for detecting and examining the relatively novel astrophysical source class of pulsar halos through planned CTA GPS observations. Using a basic phenomenological two-zone diffusion model for individual pulsar halos and their population in the Milky Way, we conduct a simulation to realistically study these entities through a comprehensive spatial-spectral likelihood analysis of simulated survey observations.

Assuming all middle-aged pulsars that exited their original PWN develop a halo, and based on a halo model set up in line with HAWC observations of the PSRJ0633+1746 halo, we estimate that around three hundred objects could produce detectable emission components in the survey. However, only about a third of these could be identified via their energy-dependent morphology, and approximately a tenth would permit strong constraints on crucial physical parameters, such as the magnitude or extent of suppressed diffusion around the pulsar. These estimates are approximately reduced by four with a model setup consistent with HAWC observations of PSRB0656+14.

Pulsar halos sustained by particle injection power in the range $10^{35-36}$ erg/s and situated a few kpc away should enable precise spectral studies from a few hundreds of GeV to tens of TeV through CTA GPS observations. The $0.1-1$ TeV band accessible to CTA presents significant potential for constraining halo transport properties. CTA is anticipated to complement HAWC and LHAASO by extending the energy coverage below 1 TeV, where the emitting particles are less influenced by energy losses and can thus explore a larger volume around the pulsar.

\textbf{Acknowledgements.} This research has made use of the CTA instrument response functions provided by the CTA Consortium and Observatory, see \href{http://www.cta-observatory.org/science/cta-performance}{http://www.cta-observatory.org/science/cta-performance} for more details. We gratefully acknowledge financial support from the agencies and organizations listed here: \href{http://www.cta-observatory.org/consortium_acknowledgments/}{https://www.cta-observatory.org/consortium\_acknowledgments/}

\bibliographystyle{JHEP}
\setlength{\bibsep}{0pt} 
\begingroup
\setstretch{1.}
{\footnotesize \bibliography{CTA_PSR_halos.bib}}
\endgroup

%
%
%

\end{document}